\documentclass[jkps,preprint,fleqn,showpacs,showkeys]{revtex4}
\usepackage{graphicx}
\usepackage{amssymb}
\usepackage{amsmath}
\usepackage{bm}
\begin{document}
\setcounter{page}{0}
\title[]{A Genetic-algorithm-based Method to Find Unitary Transformations for Any Desired Quantum Computation and Application to a One-bit Oracle Decision Problem}
\author{Jeongho \surname{Bang}}
\email{jbang@snu.ac.kr}
\affiliation{Center for Macroscopic Quantum Control \& Department of Physics and Astronomy, Seoul National University, Seoul 151-747, Korea,}
\affiliation{Department of Physics, Hanyang University, Seoul 133-791, Korea}

\author{Seokwon \surname{Yoo}}
\affiliation{Department of Physics, Hanyang University, Seoul 133-791, Korea}

\date[]{}

\begin{abstract}
We propose a genetic-algorithm-based method to find the unitary transformations for any desired quantum computation. We formulate a simple genetic algorithm by introducing the ``genetic parameter vector'' of the unitary transformations to be found. In the genetic algorithm process, all components of the genetic parameter vectors are supposed to evolve to the solution parameters of the unitary transformations. We apply our method to find the optimal unitary transformations and to generalize the corresponding quantum algorithms for a realistic problem, the one-bit oracle decision problem, or the often-called Deutsch problem. By numerical simulations, we can faithfully find the appropriate unitary transformations to solve the problem by using our method. We analyze the quantum algorithms identified by the found unitary transformations and generalize the variant models of the original Deutsch's algorithm.
\end{abstract}



\maketitle

\newcommand{\bra}[1]{\left<#1\right|}
\newcommand{\ket}[1]{\left|#1\right>}
\newcommand{\abs}[1]{\left|#1\right|}
\newcommand{\expt}[1]{\left<#1\right>}
\newcommand{\braket}[2]{\left<{#1}|{#2}\right>}
\newcommand{\commt}[2]{\left[{#1},{#2}\right]}

%-----------------------------
\section{Introduction}\label{sec:1}
%-----------------------------

Any quantum computation (QC) is implemented either implicitly or explicitly through three fundamental steps: A quantum state is initially prepared; then, the prepared state is manipulated by using a set of unitary transformations often referred to as a ``quantum algorithm.'' Finally, a measurement is performed on the output state to extract the useful information of the solution. Thus, the preparation $P$, the operation $U$, and the measurement $M$ are fundamental elements of the standard QC \cite{Nielsen00}. With this fundamental $P$-$U$-$M$ building block, one of the primary objectives in QC is to achieve the target state of the solution. However, sometimes it is very difficult or even impractical due to the lack of knowledge of the operation $U$ \cite{Bisio10}. In particular, finding the unitary transformations for any desired $U$ in QC is challenging when only limited information is available, e.g., in designing a quantum algorithm \cite{Bang08, Bang14}.

An operation $U$ can generally be described by the complete-positive trace-preserving map: $\hat{\rho}_\text{in} \rightarrow \hat{\rho}_\text{out} = \sum_k \hat{A}_k\hat{\rho}_\text{in}\hat{A}_k^\dagger$, where $\hat{\rho}_\text{in}$ is an initial state, and $\hat{A}_k$ is known as the Kraus operator, satisfying $\sum_k \hat{A}_k^\dagger\hat{A}_k = \hat{\openone}$. Such a general process of the quantum operation $U$ can also be described with an overall unitary $\hat{U}_\text{tot}$, such as $\hat{\rho}_\text{in}\otimes\ket{\cal A}\bra{\cal A} \xrightarrow{{\hat{U}_\text{tot}}, \text{Tr}_{\cal A}} \hat{\rho}_\text{out} = \text{Tr}_{\cal A}\ket{\psi_\text{out}}\bra{\psi_\text{out}}$ in a quantum system composed of a main and an extra (${\cal A}$) system, followed by a partial measurement (denoted as the partial trace `$\text{Tr}_{\cal A}$') projecting the coupled output state $\ket{\psi_\text{out}}$ on a state $\hat{\rho}_\text{out}$ \cite{Nielsen00,Audretsch07}. Here, $\ket{\cal A}$ is a state of the extra system. Thus, in a general sense, we can translate the problem into the task of finding the unitary transformation $\hat{U}_\text{tot}$ for any desired QC even in the case of the non-unitary process. Here, the extra system ${\cal A}$ and the partial measurement performed on the sub-system can arbitrarily be designed. 

Our basic idea to approach the problem, i.e., finding the (unknown) unitary transformation $\hat{U}_\text{tot}$, is to use the genetic algorithm (GA). The GA is one of the global optimization methods inspired by the biological evolution, i.e., breeding a population in which more {\em fit} individuals will have higher chances to produce their offsprings by crossing over the genetic information \cite{Holland75}. GA methods have long been used in various fields of science and engineering \cite{Forrest93,Houck95} due to their novel ability to find the optimal (unknown) solutions with limited a priori information. Thus, GA methods (or variant methods) have attracted attention lately in various applications to the quantum information and computation \cite{Mohammadi08,SaiToh14}, e.g., in laser pulse shaping \cite{Judson92,Assion98}, optimizing the measure of entanglement \cite{Munoz06}, unitary decomposition \cite{Manu12}, dynamic decoupling \cite{Quiroz13}, etc. \cite{Bang12,McDonald13,Biswas13}.

In the present paper, we propose a GA-based method to find the unitary transformations for any desired QC. We first assume that an overall QC process is decomposed into a finite series of internal unitary transformations. The only available information is a set $T$ of input-target pairs. Then, our primary problem is to find the appropriate internal unitary transformations for the given $T$. For the problem, we formulate a simple GA on a general design of QC. Here, we introduce the notion of the ``genetic parameter vector,'' which is allowed to evolve during the GA process. An initial population of the genetic parameter vectors is supposed to evolve to the corresponding unitary transformations to the desired QC. We apply our method to find the optimal internal unitary transformations and to generalize the corresponding quantum algorithm for the one-bit oracle decision problem, or the often-called Deutsch problem \cite{Deutsch85,Deutsch92}. By numerical simulations, we can faithfully find the appropriate unitary transformations to solve the problem. We analyze the quantum algorithms identified by the found unitary transformations and show that they are not exactly equal to the original Deutsch's algorithm, but correspond to its variant models.

%---------------------------------------------------------
\section{Concept \& method}\label{sec:2}
%---------------------------------------------------------

We first need to specify the problem more precisely. As described in the previous section, the preparation $P$, the operation $U$, and the measurement $M$ are basic elements of the standard QC. The operation $U$ can generally be described by an overall unitary $\hat{U}_\text{tot}$ with an extra system and an arbitrarily chosen partial measurement. In practical QC, however, a finite number of internal unitary transformations are usually designed to implement $\hat{U}_\text{tot}$ of the desired QC, taking into account the proper minimum cost of the realization \cite{Cybenko01,Daskin11}. Thus, we assume that $\hat{U}_\text{tot}$ is decomposed into a finite $N_u$ series of internal unitary transformations $\hat{U}_j$ ($j=1,2,\ldots,N_u$) whose exact forms are also {\em yet to be known}. Here, we assume further that the only available information is a set $T$ of input-target pairs. The input $x$ is {\em classical information}, because we -- a {\em classical} supervisor or a {\em classical} algorithm designer -- must perceive it. The information $x$ is often provided as a functional form and is usually encoded into $P$ or $U$ in QC whereas the target is a desired output {\em quantum state} $\ket{\tau_x}$ for the given $x$. In the circumstance, the main problem dealt with here is to find a set of appropriate internal unitary transformations $\hat{U}_1, \hat{U}_2, \ldots, \hat{U}_{N_u}$ for the given $T$.

Here, we briefly note that any essential part of using the information $x$ is always involved in QC. In a typical scheme of QC, the input $x$ is usually encoded in an ancillary system or a relative phase of the internally evolving quantum state. Such an encoding for the reference of $x$ can successfully be performed by using a specific unitary operation, the so-called quantum oracle. We clarify that such a non-trivial operation would be involved in each separated unitary transformation $\hat{U}_j$ ($j=1,2,\ldots,N_u$) or in one of them.

\begin{figure}[t]
\centering
\includegraphics[width=0.5\textwidth]{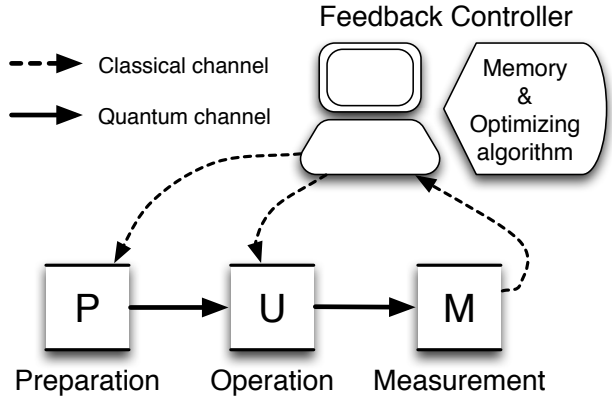}
\caption{General design of our method. This architecture is borrowed from a general model for a quantum computation process ($P$-$U$-$M$) assisted by a feedback system ($F$).}
\label{fig:scheme}
\end{figure}

We then introduce another element, called the feedback controller $F$, which contains an optimizing algorithm and a finite size of the classical memory. The basic architecture of our method is, thus, borrowed from a general model of QC assisted by a feedback system (See Fig.~\ref{fig:scheme}). Here, we note that $F$ is {\em classical} in the sense that it mainly deals with classical information, e.g., control parameters of $U$ and measurement outcomes from $M$. This information is communicated through the classical channel.

In such a basic design, the optimizing algorithm in $F$ is particularly important because it is directly connected with the efficiency, accuracy, and other performances of the method. In this work, we employ the genetic algorithm (GA), which is one of the widely-used global optimization methods \cite{Houck95}. Typically, the GA runs as follows: ($i$) One prepares a population as the set of candidate solutions, ($ii$) selects several candidates to generate their offsprings, and then ($iii$) reconstructs a new population with newly-generated offsprings. Using fitness criteria, the candidates can breed their offsprings. The above processes ($i$)-($iii$) are continued to meet a certain `termination condition.' In formulating a GA, the most fundamental and important issue is to represent the genetic information of the candidate solutions \cite{Rothlauf06}. The question of how we define a certain condition to terminate the process is also one of the important issues and that problem remains open \cite{Aytug96}. With these issues, we now formulate a simple GA.

Before starting, we briefly note here that our formulation refers to the standard (or simple) GA model initially introduced and studied by John Holland \cite{Holland75} because the main purpose of this work is to provide a basic framework rather than to develop it. Most existing theories and applications were also build upon this standard GA model, although some remarkable theoretical advances were primarily built on its variant models \cite{Whitley94}. 

{\em Population preparation.} -- First, we should prepare a population as the set of candidate solutions. From now on, we call a single item of the candidate solution an ``individual'' (just as in a standard GA model \cite{Houck95}). In our case, $\hat{U}_\text{tot}$ corresponds to an individual. Noting that $\hat{U}_\text{tot}$ consists of the internal unitary transformations $\hat{U}_j$ ($j=1,2,\ldots,N_u$), we represent a population as the number $N_\text{pop}$ of unitary sets, 
\begin{eqnarray}
\left\{\hat{U}_1^{(n)}, \hat{U}_2^{(n)}, \ldots, \hat{U}_{N_u}^{(n)} \right\}_{n=1}^{N_\text{pop}},
\label{eq:pop_u}
\end{eqnarray}
which is initially prepared at random. 

We parametrize the internal unitary transformation $\hat{U}_j$ in $d$-dimensional Hilbert space as
\begin{eqnarray}
\hat{U}_j(\mathbf{p}_j)=e^{-i\mathbf{p}_j^T\boldsymbol{\sigma}},
\label{eq:u_op}
\end{eqnarray}
where $\mathbf{p}_j=(p_1, p_2, \ldots , p_{d^2-1})_j^T$ is a real vector in $(d^2-1)$-dimensional real vector space $\mathbb{R}^{d^2-1}$, and $\boldsymbol{\sigma} = (\hat{\sigma}_1, \hat{\sigma}_2,\ldots,\hat{\sigma}_{d^2-1})^T$ is a vector whose components are SU($d$) group generators \cite{Hioe81,Son04}. Note that, in our method, a component $p_k \in [-\pi, \pi]$ ($k = 1, 2, \ldots, d^2-1$) would be represented by a genetic form in order to evolve the internal unitary transformations (as detailed below). Our method is, in principle, applicable to a real experiment, as $p_k$'s can be directly matched to the control parameters, e.g., beam-splitter and phase-shifter alignments in linear optical system \cite{Reck94} or radio-frequency (rf) pulse sequences in nuclear magnetic resonance (NMR) system \cite{Lee00}.

\begin{figure}[t]
\centering
\includegraphics[width=0.5\textwidth]{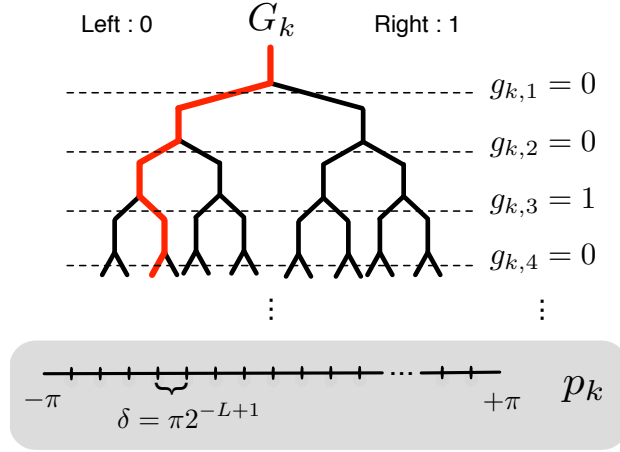}
\caption{(Color online) A chromosome $G_k$ is characterized as left or right branches (genes) on a binary tree. For example, ``$G_k = 0010\cdots$'' corresponds to ``left(0)-left(0)-right(1)-left(0)-$\cdots$'' (denoted by a red line). Note that any flow finally arrives at a certain value of $p_k$ discretized to $2^L$ points with spacing $\delta = \pi 2^{-L+1}$ in $(-\pi, \pi)$ [See the mapping Eq.~(\ref{eq:n_param})].}
\label{fig:g_path}
\end{figure}

{\em Genetic representation.} -- We give here the genetic representation of the real vector $\mathbf{p}$. The usual way is to take a finite, say $L$, length of the binary strings ($0$'s and $1$'s). Following this, we define $G_k$ (``chromosome'') as a binary $L$ sequence: $G_k = g_{k,1} g_{k,2} \ldots g_{k,L}$, where $g_{k,l} \in \{0,1\}$ (``gene''), and $L$ is a {\em depth constant}. Note that the accuracy of the found unitary transformations depends on the depth constant $L$ because the number of digits to represent a solution parameter increases with increasing $L$. However, we also note that the run-time of the GA process becomes longer for larger $L$, because the possible representation of the solutions, i.e., the size of the search space, also increases. 

For the sake of the convenience, we visualize a chromosome $G_k$ in terms of the left(`$0$') or right(`$1$') branches on a binary tree, as depicted in Fig.~\ref{fig:g_path}. Here, any flow is seen to finally arrive at a certain value of $p_k$ discretized to $2^L$ points with spacing $\delta = \pi 2^{-L+1}$ from $-\pi$ to $\pi$. By observing this, we can easily derive the mapping between $G_k$ and $p_k$ as
\begin{eqnarray}
G_k \mapsto p_k = 2\pi \left( \sum_{l=1}^{L} \frac{(-1)^{g_{k,l}\oplus1}}{2^{l}} \right),
\label{eq:n_param}
\end{eqnarray}
where `$\oplus$' denotes the modulo-$2$ addition. We, thus, consider a real vector $\mathbf{G}=(G_1, G_2, \ldots G_{d^2-1})^T$, whose components are given as the chromosomes $G_k$. We call this vector $\mathbf{G}$ the ``genetic parameter vector'' (which is an approximation of the real vector $\mathbf{p} \in \mathbb{R}^{d^2-1}$). As we can make a one-to-one correspondence between the genetic parameter vectors $\mathbf{G}_j$ and the internal unitary transformations $\hat{U}_j$ for all $j=1,\ldots,N_u$, the population as in Eq.~(\ref{eq:pop_u}) can be represented in terms of the number $N_\text{pop}$ of the genetic parameter vector sets as
\begin{eqnarray}
\left\{\mathbf{G}_1^{(n)}, \mathbf{G}_2^{(n)}, \ldots, \mathbf{G}_{N_u}^{(n)} \right\}_{n=1}^{N_\text{pop}}.
\label{eq:pop_gpv}
\end{eqnarray}

{\em Selection.} -- Selection is a step in which the particular individuals are chosen to breed. Only the selected individuals have the opportunity to transfer their genetic information. The selection is done based on the ``fitness'' which quantifies how {\em fit} the individual is for the given circumstance. In our case, the fitness $\xi_n$ of the $n^\text{th}$ individual is defined as the mean fidelity, 
\begin{eqnarray}
\xi_n = \frac{1}{N_{x\text{-}\tau}} \sum_{x \in T} \abs{\bra{\tau_x}\hat{U}_\text{tot}^{(n)} \ket{\psi_\text{in}}}^2, 
\label{eq:fidel}
\end{eqnarray}
where $\hat{U}_\text{tot}^{(n)}$ corresponds to the $n^\text{th}$ individual, $\ket{\psi_\text{in}}$ is the initial state generated in $P$, and $\ket{\tau_x}$ is the state of the target for the given $x$. Here, the summation $\sum_{x \in T}$ is done for a finite $N_{x\text{-}\tau}$ of the input-target pairs $\left( x,\ket{\tau_x} \right)$ in $T$ (i.e., $N_{x\text{-}\tau}=\abs{T}$). The maximum value of the fitness, i.e., $\xi_n =1$, implies that $\hat{U}_\text{tot}^{(n)}$ is the perfect for the desired QC while it is incomplete when $\xi_n < 1$ .

Therefore, the strategy for the selection in GA is that the individuals corresponding to better solutions are more likely to be selected; the high-fitness individuals have higher probability to be selected. In our method, the probability $P(n)$ that an $n^\text{th}$ individual is chosen is 
\begin{eqnarray}
P(n) \propto e^{- \frac{\ln{N_\textrm{pop}}}{N_\textrm{pop}-1} (n-1)},
\label{eq:sel_p}
\end{eqnarray}
where we assume that the individuals are sorted by $\xi_1 \ge \xi_2 \ge \ldots \ge \xi_{N_\textrm{pop}}$ and that $\sum_{n=1}^{N_\text{pop}}P(n)=1$. Here, $P(N_\text{pop})=\frac{1}{N_\text{pop}}P(1)$.

\begin{figure}[t]
\centering
\includegraphics[width=0.6\textwidth]{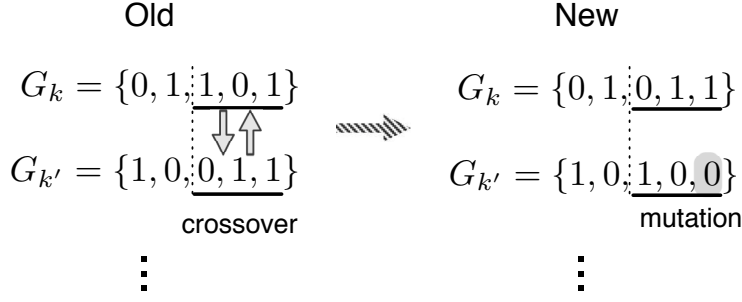}
\caption{We present a process of crossover and mutation. For simplicity, we let $L=5$. We first chose two (old) genes $G_{j,k}^{(n_1)}$ and $G_{j,k}^{(n_2)}$ by using the selection probability in Eq.~(\ref{eq:sel_p}), where $n_1 \neq n_2 \in [1, N_\text{pop}]$ ($j = 1,2,\ldots,N_u$, and $k=1,2,\ldots,d^2-1$). Any segments of the binary strings (``$101$'' in $G_{j,k}^{(n_1)}$ and ``$011$'' in $G_{j,k}^{(n_2)}$) are then exchanged to generate new ones ${G'}_{j,k}^{(n_1)}$ and ${G'}_{j,k}^{(n_2)}$. By mutation, a binary number `$1$' was flipped to `$0$' in the newly-generated ${G'}_{j,k}^{(n_2)}$.}
\label{fig:crossover}
\end{figure}

{\em Crossover and mutation.} -- Crossover and mutation are known to be the main genetic operations to evolve the population. By crossover, new individuals can be generated. In most case, the segments of the parents' genes are transferred to their offsprings. One typical method is to exchange some parts of the binary strings. In our case, a genetic parameter vector is newly generated by merging the genes in the two selected vectors, as illustrated in Fig.~\ref{fig:crossover}. Thus, we can make progress in the population by renewing all $N_\text{pop}$ sets of the genetic parameter vectors.

In mutation, some genetic information is self-generated or transformed without the crossover. It can be applied by changing an arbitrary bit string from the original one. The purpose of mutation is usually to improve the diversity or to extend the solution space \cite{Houck95}. We can realize such an operation by flipping a string in a gene $G_j$ (See also Fig.~\ref{fig:crossover}). 

{\em Termination condition.} -- The GA process would be terminated when a relevant condition is met. This condition is called the termination condition. The most easily and frequently used termination condition is to fix the maximum number of generations, taking into account the computational resources, e.g., the memory size or the scale of the problem. Another way involves the convergence of individuals; namely, if the improvement in the fitnesses becomes smaller than a threshold value, say $h$, then the process is terminated due to the lack of improvements. We consider the latter type here. 

To construct the termination condition, we consider the ``fitness fluctuation'' as the standard deviation: 
\begin{eqnarray}
{\Delta \xi} = \left(\frac{1}{N_\text{pop}}\sum_{n=1}^{N_\text{pop}}\xi_n^2 - \overline{\xi}^2\right)^\frac{1}{2}, 
\end{eqnarray}
where $\overline{\xi}=\frac{1}{N_\text{pop}}\sum_{n=1}^{N_\text{pop}}\xi_n$ is the mean fitness over the population. Then, we define the termination condition as follows: First, we set a constant $h$ to be very small. We evaluate ${\Delta\xi}$ and compare it with the predetermined value $h$ at every generation step of the GA process. In those cases where ${\Delta \xi}$ is larger than $h$, the process goes on; however, if we meet ${\Delta \xi} < h$, the process is terminated. 

We indicate here that we should take into account the rounding-off error due to the finite $L$. The rounding-off error would approximately be proportional to $O(d^2 N_u \delta)$, where $\delta = \pi 2^{-L+1}$ (See Fig.~\ref{fig:g_path}). Note that, if the dimension $d$ of Hilbert space is not too large and $\hat{U}_\text{tot}$ consists of a reasonable number $N_u$ of internal unitary transformations, the rounding-off error can be made vanishingly small by choosing a sufficiently large $L$.

%------------------------------------------
\section{Application to the one-bit oracle decision problem}\label{sec:3}
%------------------------------------------

\begin{figure}[t]
\centering
\includegraphics[width=0.5\textwidth]{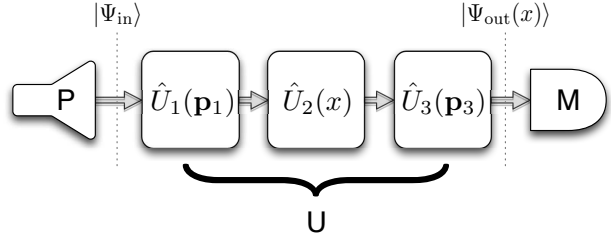}
\caption{Quantum circuit for the one-bit oracle decision problem. $\hat{U}_2$ is the unitary of the oracle operation. The other two, $\hat{U}_1$ and $\hat{U}_3$, allow us to generate the desired output $\ket{\psi_\text{out}(x)}$ as in Eq.~(\ref{eq:out_st}) by only one evaluation of $x$.}
\label{fig:qc_da}
\end{figure}

We apply our method to a realistic problem known as the one-bit oracle decision problem, or often called the ``Deutsch'' problem. This problem is to decide if an arbitrary one-bit Boolean function $x:\{0,1\} \rightarrow \{0,1\}$ is constant, i.e., $x(0)=x(1)$, or balanced, i.e., $x(0) \neq x(1)$. Classically, the function $x$ should be evaluated twice for $0$ and $1$ to solve the problem. On the other hand, QC enables us to identify the function $x$ by using only one evaluation \cite{Deutsch85,Deutsch92}. The quantum circuit for such an algorithm is presented in Fig.~$\ref{fig:qc_da}$. In the circuit, the unitary transformation $\hat{U}_2$ corresponds to the oracle operation defined by
\begin{eqnarray}
\hat{U}_2\ket{k} = e^{i\pi x(k)}\ket{k},~k \in \{0,1\},
\end{eqnarray}
where $\ket{0}$ and $\ket{1}$ are computational bases in a qubit system. Such a form of the oracle is widely used in QC \cite{Kashefi02}. The other two unitary transformations $\hat{U}_1$ and $\hat{U}_3$ change the incoming states to the superposed states so that we get the final output state as
\begin{eqnarray}
\ket{\psi_\textrm{out}(x)} =
\left\{
	\begin{array}{ll}
	\ket{m_0} &~\text{if $x$ is constant}, \\
	\ket{m_1} &~\text{if $x$ is balanced},
	\end{array}
\right.
\label{eq:out_st}
\end{eqnarray}
where $\ket{m_0}$ and $\ket{m_1}$ are arbitrary qubit states orthogonal to each other, i.e., $|\braket{m_0}{m_1}|^2=0$. We then identify the given function $x$ by performing the (von-Neumann) measurement $\hat{M} = \sum_{l=0,1}(-1)^l \ket{m_l}\bra{m_l}$. Here, if $\ket{m_0}$ is measured, then $x$ is a `constant' function; otherwise, $x$ is a `balanced' function.

We can easily understand how this algorithm works. First, $\hat{U}_1$ distributes the initially prepared state $\ket{\Psi_\textrm{in}}$ to an arbitrarily superposition of $\ket{0}$ and $\ket{1}$; then, $\hat{U}_2$ acts {\em only once} on the distributed state. Finally, $\hat{U}_3$ leads the incoming state to the corresponding output $\ket{\psi_\text{out}(x)}$ for decision. The key feature of the algorithm is the ``quantum parallelism,'' by which all the values of $0$ and $1$ are {\em simultaneously} evaluated in the form of their superposition. Therefore, finding appropriate unitary transformations $\hat{U}_1$ and $\hat{U}_3$ to maximize the quantum parallelism is important. In the original Deutsch's algorithm, $\hat{U}_1$ and $\hat{U}_3$ are given as the Hadamard operation $\hat{H}$, which transforms $\ket{0}$ and $\ket{1}$ into an equally superposed state, such as $\hat{H}\ket{0}\rightarrow\left(\ket{0}+\ket{1}\right)/{\sqrt{2}}$ and $\hat{H}\ket{1}\rightarrow\left(\ket{0}-\ket{1}\right)/{\sqrt{2}}$, and the final measurement is $\hat{M} = \ket{0}\bra{0} - \ket{1}\bra{1}$. Here, we note that $\hat{U}_2$ is also very important, as it is employed to encode the input $x$ in our method.

Thus, we apply our method to find other explicit forms of $\hat{U}_1$ and $\hat{U}_3$ and to generalize the Deutsch's algorithm. First, let us consider an input-target set
\begin{eqnarray}
T=\left\{\left(x_i=c, \ket{\tau_c}=\ket{0}\right), \left(x_i=b, \ket{\tau_b}=\ket{1}\right)\right\},
\end{eqnarray}
where the input $x_i$ ($i=1,2,3,4$) is one of the four possible Boolean functions, and `$c$' and `$b$' stand for the constant and the balanced function, respectively. We then consider a decomposition of $\hat{U}_\text{tot}$ such that
\begin{eqnarray}
\hat{U}_\text{tot} = \hat{U}_3(\mathbf{p}_3) \hat{U}_2(x_i) \hat{U}_1(\mathbf{p}_1),
\label{eq:uop}
\end{eqnarray}
where $\hat{U}_2$ is a part of encoding the given function $x_i \in \{c, b\}$, and the other two $\hat{U}_1$ and $\hat{U}_3$ are controllable single-qubit unitary transformations. By `controllable,' we mean here that $\hat{U}(\mathbf{p}_j)$ can be controlled by adjusting $\mathbf{p}_j$ ($j=1,3$). The preparation $P$ generates $\ket{\psi_\text{in}}$, and a measurement $M$ is chosen for the decision of $x$. The feedback $F$ is responsible for the GA process. From Eq.~(\ref{eq:fidel}), we represent the fitness $\xi_n$ of any $n^\text{th}$ individual $\hat{U}_\text{tot}^{(n)}$ as
\begin{eqnarray}
\xi_n = \frac{f_c + f_b}{2},
\end{eqnarray}
where $f_\kappa = |\bra{\tau_\kappa}\hat{U}_3^{(n)} \hat{U}_2(x_i=\kappa) \hat{U}_1^{(n)}\ket{\psi_\text{in}}|^2$ ($\kappa = b, c$).

\begin{figure}[t]
\centering
\includegraphics[angle=270,width=0.35\textwidth]{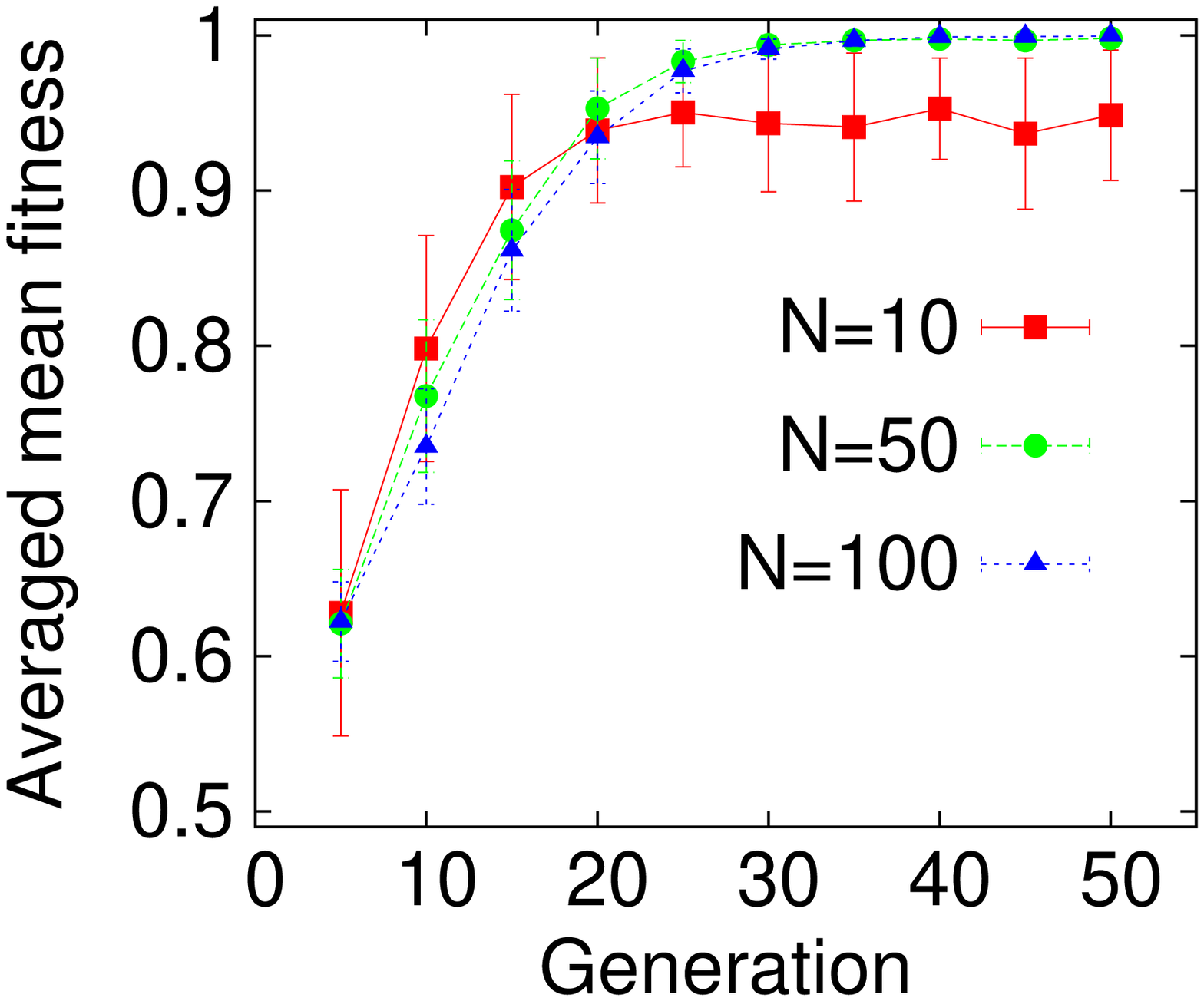}
\includegraphics[angle=270,width=0.35\textwidth]{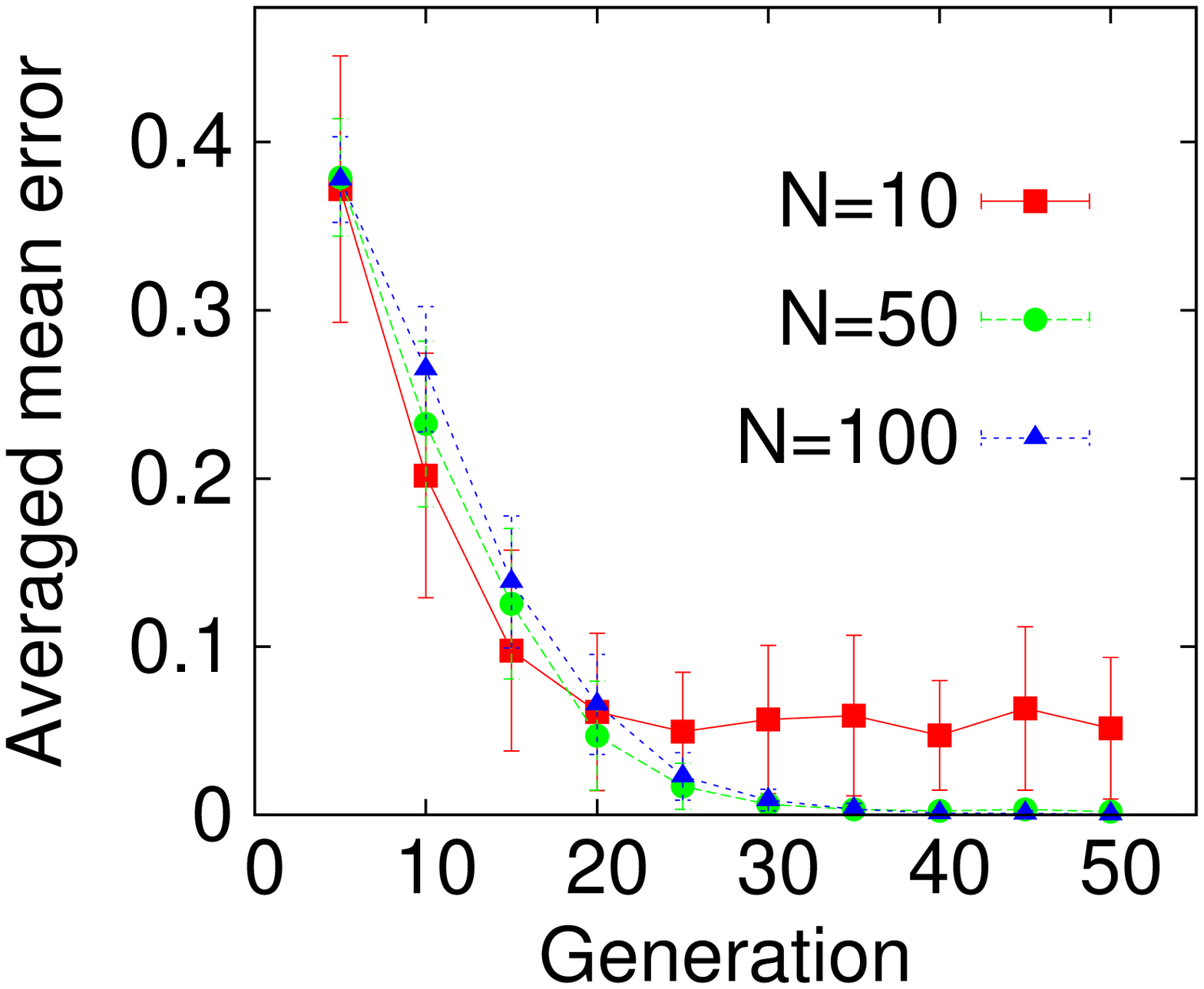}
\caption{(Color online) We plot (left) the average mean fitnesses $\overline{\xi}_\text{av}=\sum_{n=1}^{1000}\overline{\xi}_i$ and (right) the averaged mean error $\overline{\epsilon}_\text{av} = 1-\overline{\xi}_\text{av}$. We consider three population sizes: $N_\text{pop}=10$, $50$, and $100$. Each point is obtained by averaging over $1000$ simulations, and the error bar is the standard deviation. It is directly seen that $\overline{\xi}_\text{av}$ is increased or $\overline{\epsilon}_\text{av}$ is decreased, as the generation proceeds.}
\label{grp:ga_p}
\end{figure}

Based on the above settings, the numerical simulations are carried out. In the simulations, we prepare $N_\text{pop}$ individuals as a population. Thus, we have $N_\text{pop}$ sets of $3$-dimensional genetic parameter vectors as 
\begin{eqnarray}
\left\{ \mathbf{G}_1^{(n)}, \mathbf{G}_3^{(n)} \right\}_{n=1}^{N_\text{pop}}.
\end{eqnarray}
Here, we consider three cases: $N_\text{pop}=10$, $50$, and $100$. We let $L=15$ to ignore the rounding-off error. For the sake of simplicity, we take $\ket{\psi_\text{in}}=\ket{0}$, and the mutation is not considered. In Fig. $\ref{grp:ga_p}$, we give the mean fitnesses $\overline{\xi}_\text{av}$ averaged over $1000$ simulations. The error bars are the standard deviations. As directly seen in Fig.~\ref{grp:ga_p}, $\overline{\xi}_\text{av}$ is increased, or equivalently, $\overline{\epsilon}_\text{av}=1-\overline{\xi}_\text{av}$ is decreased. In particular, we can see that $\overline{\xi}_\text{av} \rightarrow 1$ (or $\overline{\epsilon}_\text{av} \rightarrow 0$) fora large $N_\text{pop}$. We note here that, for a given problem size $D$, a large number $N_\text{pop}$ of populations is usually needed in evolutionary optimization. For example, choosing $N_\text{pop} \simeq 5D \sim 10D$ is appropriate (See Ref.~\cite{Storn97}). In our case, the problem size $D$ is given as $2(d^2-1)=6$, which is the total number of control parameters in $\hat{U}_j(p_1,p_2,p_3)$ ($j=1,3$). However, we can still achieve a high accuracy even for a relatively small population size. Actually, when $N_\text{pop}=10$, $\overline{\xi}_\text{av}$ is as high as $0.949 \pm 0.042$ at the $50^\text{th}$ generation step. 

We here find that the identified $\hat{U}_\text{tot}$ will not be exactly equal to the original Deutsch's algorithm, but will correspond to one variant model of it, with $\hat{U}_{1} \neq \hat{U}_{3} \neq \hat{H}$. To see this, it is convenient to rewrite the single-qubit unitary $\hat{U}_j$ ($j=1,3$) in the following (geometric) form:
\begin{eqnarray}
\hat{U}_j(\mathbf{p}_j) = e^{-i \mathbf{p}_j^T \boldsymbol{\sigma}} = \cos{\Theta_j}\hat{\openone} - i \sin{\Theta_j}\left( \mathbf{n}_j^T \boldsymbol{\sigma} \right),
\end{eqnarray}
where $\boldsymbol{\sigma}=(\hat{\sigma}_1, \hat{\sigma}_2, \hat{\sigma}_3)^T$ is the vector of Pauli operators, $\Theta_j$ is given in terms of the Euclidean norm of $\mathbf{p}_j$, i.e. $\Theta_j = \lVert \mathbf{p_j} \rVert = (\mathbf{p}_j^T \mathbf{p}_j)^\frac{1}{2}$, and $\mathbf{n}_j=\frac{\mathbf{p}_j}{\lVert \mathbf{p_j} \rVert}$ is the normalized vector. Any pure quantum state is characterized as a point on the surface of a unit sphere, called a ``Bloch sphere,'' and $\hat{U}_j$ rotates a pure state, a point on the Bloch sphere, by the angle $2\Theta_j$ around the axis $\mathbf{n}_j$ \cite{Tian04}. For example, in the case of the original Deutsch's algorithm, the Hadamard operation $\hat{H}$ (corresponding to $\hat{U}_1$ and $\hat{U}_3$) is a $\pi$-rotation about the axis $\mathbf{n}=({1}/{\sqrt{2}}, 0, {1}/{\sqrt{2}})^T$. 

On the basis of the above description, we can generally describe how the identified algorithm $\hat{U}_\text{tot,opt}=\hat{U}_3(\mathbf{p}_{3,\text{opt}})\hat{U}_2(x_i)\hat{U}_1(\mathbf{p}_{1,\text{opt}})$ works: First, $\hat{U}_1$ rotates the initial state $\ket{\psi_\text{in}}$ to
\begin{eqnarray}
\hat{U}_1\ket{\psi_\text{in}} = \alpha\ket{0} + e^{i \phi}\sqrt{1-\alpha^2}\ket{1},
\label{eq:necessary_c}
\end{eqnarray}
where $\phi$ is the phase factor. Then, the oracle $\hat{U}_2$ flips the phase $\phi$ to $\phi+\pi$ if $x_i=b$ and leaves it unchanged if $x_i=c$. The last unitary transformation $\hat{U}_3$ rotates the state to the corresponding output (e.g., $\ket{\psi_\text{out}(x_i = c)} = \ket{0}$ and $\ket{\psi_\text{out}(x_i = b)} = \ket{1}$ in the case of the original Deutsch's algorithm). Here, we provide the necessary condition for the Deutsch's algorithm: {\em The state of Eq.~(\ref{eq:necessary_c}) should be on the equator of the Bloch sphere}, or equivalently, {\em $\alpha$ should be equal to $1/\sqrt{2}$} (See Appendix~\ref{appendix:variants}). In Fig.~\ref{grp:variants}, we plot $1000$ data points ($\alpha$, $\phi$) characterized by the found $\hat{U}_1(\mathbf{p}_{1,\text{opt}})$ in polar coordinates. Here, we have $\overline{\alpha}\simeq 0.708 \simeq 1/\sqrt{2}$ and ${\Delta\alpha}\simeq 0.004$, where $\overline{\alpha}$ and ${\Delta\alpha}$ denote the average and the standard deviation, respectively.

\begin{figure}[t]
\centering
\includegraphics[angle=270,width=0.65\textwidth]{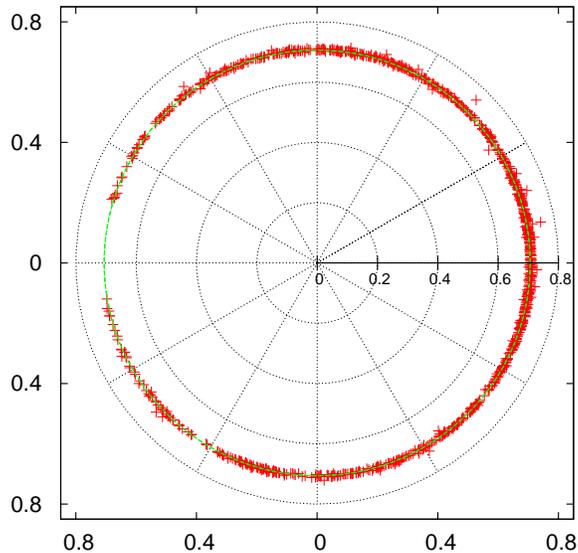}
\caption{(Color online) We plot $\alpha$ and $\phi$ in Eq.~(\ref{eq:necessary_c}) in polar coordinates ($\alpha$, $\phi$). Each data point is made by using the found $\hat{U}_1(\mathbf{p}_{1,\mathbf{opt}})$ and averaged over $1000$ simulation results. As analyzed, the values of $\alpha$ are located on $\simeq 1/\sqrt{2}$ (green circular line) whereas $\phi$ has an arbitrary value; actually, we have $\overline{\alpha}\simeq 0.708 \simeq 1/\sqrt{2}$ and ${\Delta\alpha}\simeq 0.004$, where $\overline{\alpha}$ and ${\Delta\alpha}$ are the average and the standard deviation, respectively.}
\label{grp:variants}
\end{figure}

\begin{figure}[t]
\centering
\includegraphics[angle=270,width=0.35\textwidth]{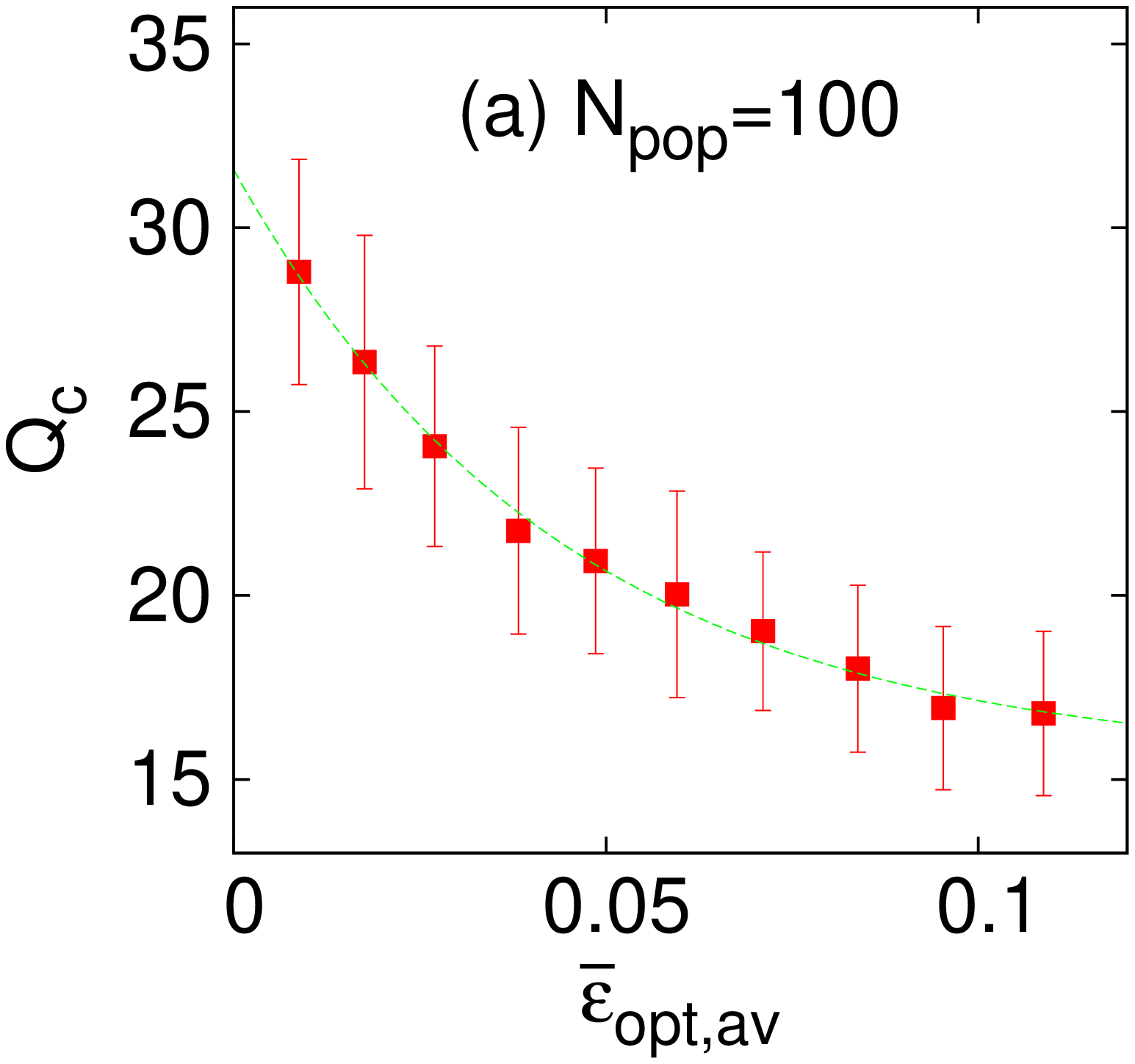}
\includegraphics[angle=270,width=0.35\textwidth]{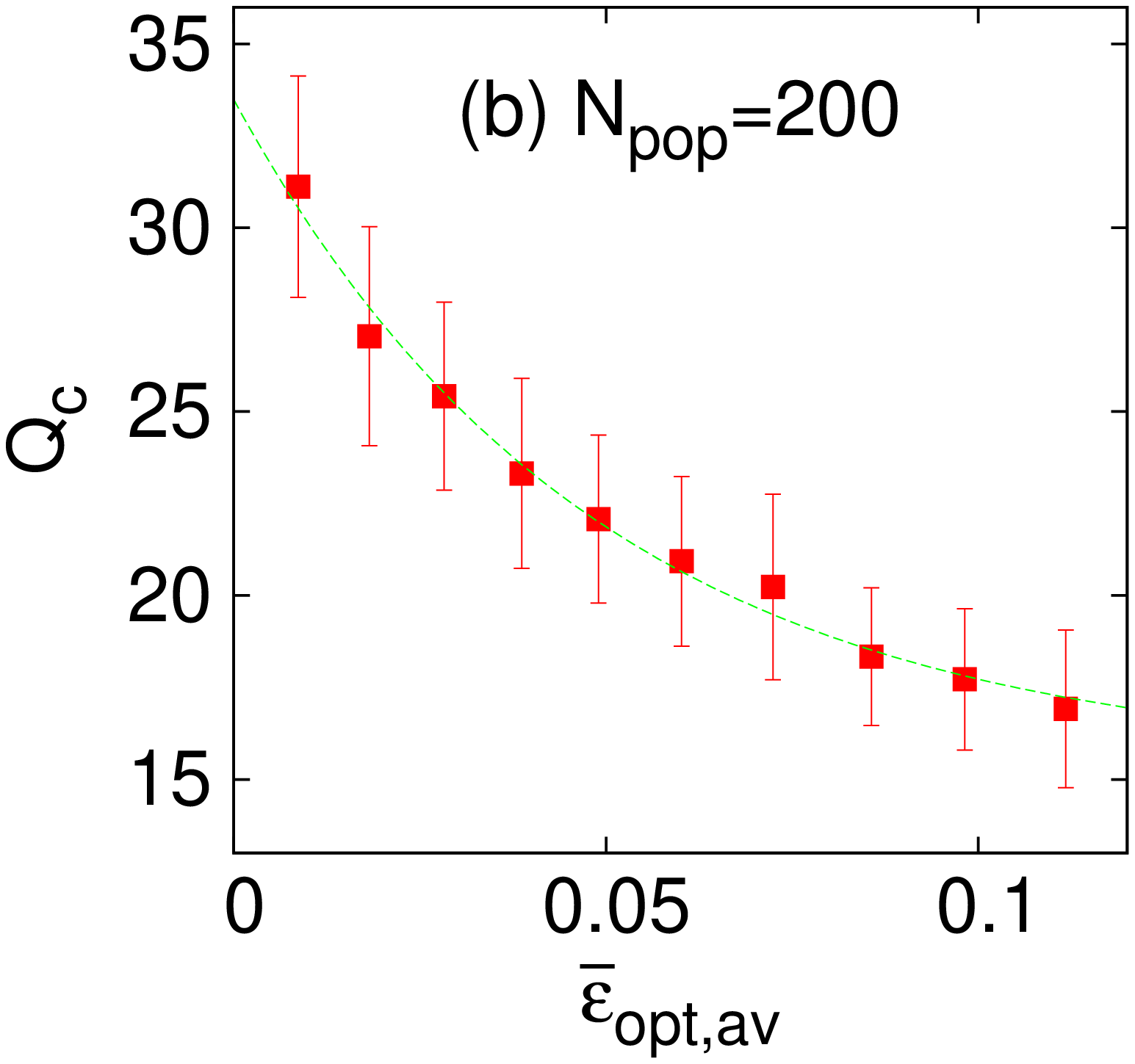} \\
\includegraphics[angle=270,width=0.35\textwidth]{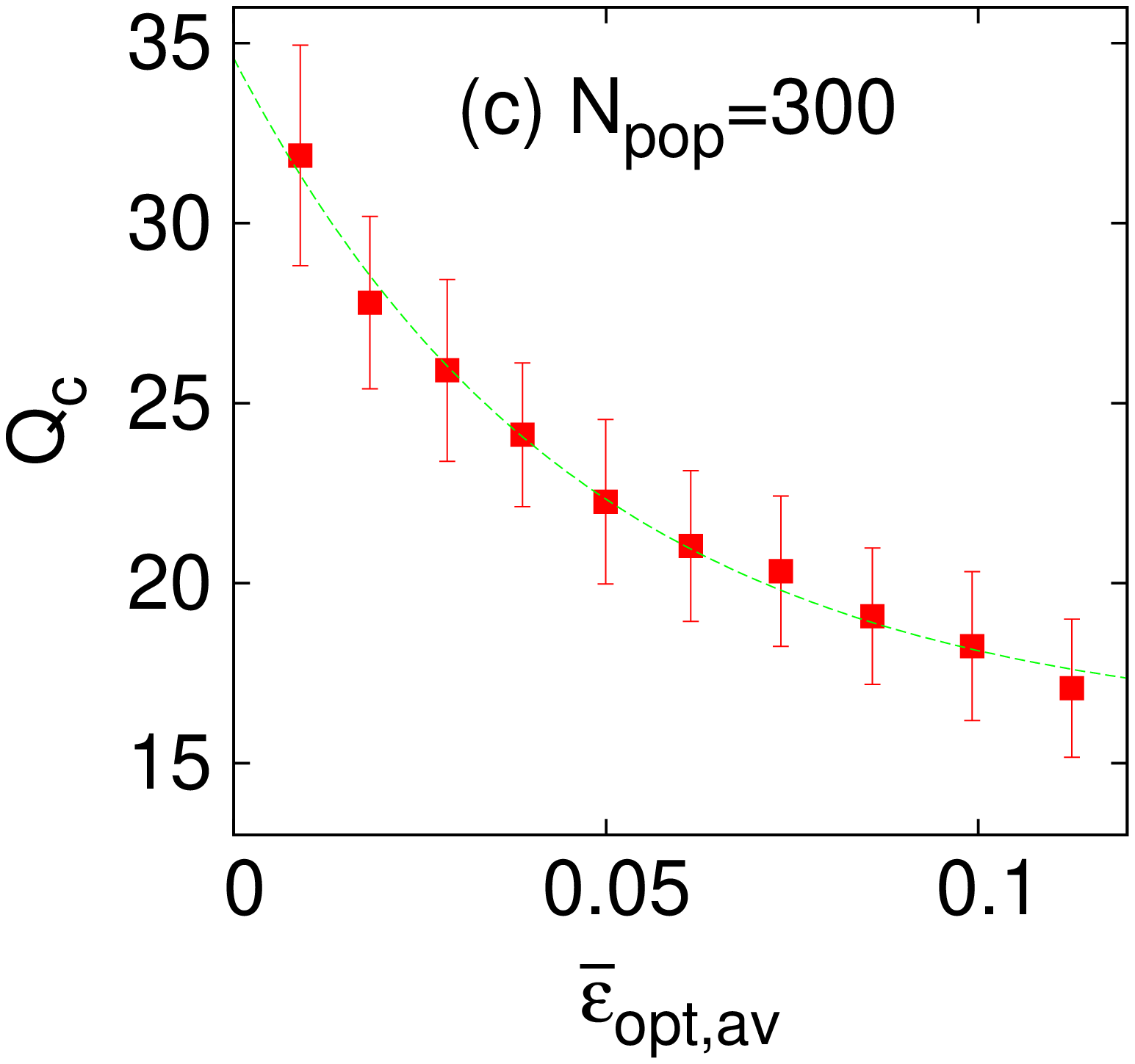}
\includegraphics[angle=270,width=0.35\textwidth]{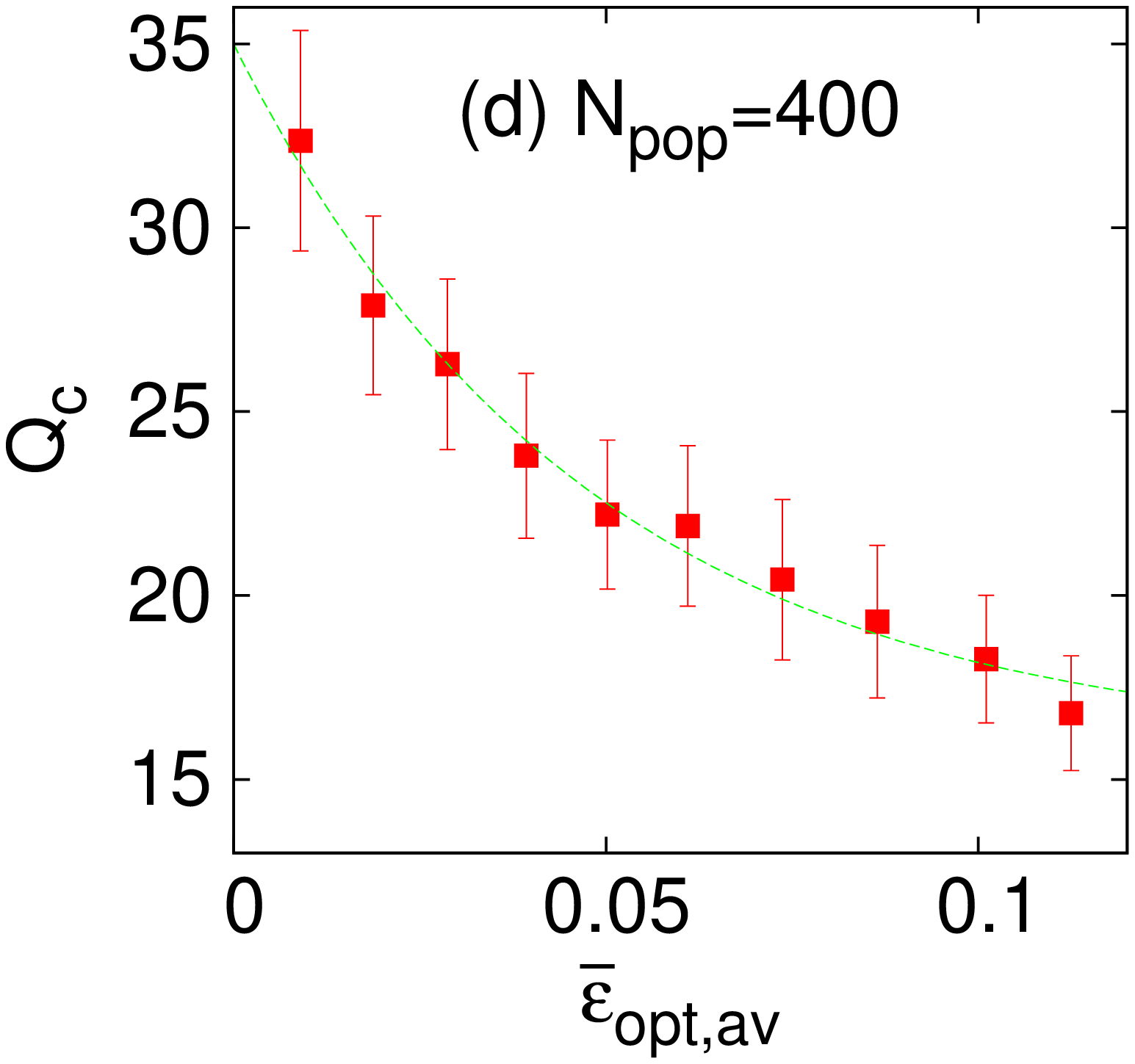}
\caption{(Color online) $Q_c$ versus $\overline{\epsilon}_\text{opt,av}$ graphs are given for (a) $N_\text{pop}=100$, (b) $200$, (c) $300$, and (d) $400$. The data are well fitted to $Q_c = \alpha e^{-\beta \overline{\epsilon}} + \gamma$. Fitting parameters $\alpha$, $\beta$, and $\gamma$ are listed in table~\ref{tab:nc}.}
\label{grp:err_nc}
\end{figure}

We perform numerical simulations for further analysis. In particular, we investigate the relation between the required steps $Q_c$ of the generations to complete the GA process and the accuracy of the identified algorithm $\hat{U}_\text{tot,opt}$ after the completion. To do this, we evaluate the optimal mean errors $\overline{\epsilon}_\text{opt} = 1 - \overline{\xi}_\text{opt}$ of the identified $\hat{U}_\text{tot,opt}$ and find $Q_c$ in each simulation. The simulation is performed $1000$ times for each case: $N_\text{pop}=100$, $200$, $300$, and $400$. We let $L=15$, as above. In Fig.~\ref{grp:err_nc}, we give graphs of $Q_c$ versus $\overline{\epsilon}_\text{opt,av}= \sum_{i=1}^{1000}\overline{\epsilon}_{\text{opt},i}$. Note that the GA process is terminated within $\simeq 35$ for all cases of $N_\text{pop}$. We assume here an {\em exponential} dependence of the overall run-time ($Q_c$ in our case) on the required accuracy, which is typically found in evolutionary optimization (See Refs.~\cite{Bergh04} and \cite{Chu11}, or references therein). With this assumption, we find that the data are well fitted to a function $Q_c = \alpha e^{-\beta \overline{\epsilon}} + \gamma$. The detailed fitting parameters, $\alpha$, $\beta$, and $\gamma$, are given in table~\ref{tab:nc}. Note that the estimated parameters for each $N_\text{pop}$ are all consistent (within their error ranges).

\begin{table}[t]
	\begin{tabular}{cccc} 
		\toprule
		$N_\text{pop}$ & $\alpha$ & $\beta$ & $\gamma$ \\
		\colrule
		$100$ & $16.13 \pm 0.44$ & $22.58 \pm 2.25$ & $15.46 \pm 0.53$ \\
		$200$ & $18.06 \pm 0.73$ & $21.63 \pm 2.98$ & $15.43 \pm 0.92$ \\
		$300$ & $18.65 \pm 0.64$ & $21.38 \pm 2.65$ & $15.92 \pm 0.79$ \\
		$400$ & $19.13 \pm 0.93$ & $21.09 \pm 3.07$ & $15.85 \pm 1.15$ \\
		\botrule
	\end{tabular}
	\caption{Fitting parameters $\alpha$, $\beta$, and $\gamma$ for the data in Fig.~\ref{grp:err_nc}, with $g_c = \alpha e^{-\beta \overline{\epsilon}} + \gamma$.}
	\label{tab:nc}
\end{table}

%------------------------------
\section{Summary}
%------------------------------

We have proposed a naive method based on the genetic algorithm (GA) to find the unitary transformations for any desired quantum computation (QC). To specify the problem, we assumed that an overall unitary for QC could be decomposed to a finite series of internal unitary transformations. Here, we also assumed that the only available information would be a set of input-target pairs. With these assumptions, the problem was to find the appropriate internal unitary transformations for the given input-target set. Thus, we formulated the simple GA by introducing the notion of the genetic parameter vector. The genetic parameter vectors of the internal unitary transformations were allowed to evolve in the GA process. We argued that the presented method can, in principle, be applied to a real experiment with the current technology. 

We then applied our method to find the optimal unitary transformations and to generalize the corresponding quantum algorithm for a realistic QC problem, known as the one-bit oracle decision problem, or the often-called Deutsch problem. By numerical simulations, we showed that the appropriate unitary transformations to solve the problem can faithfully be found in our method. We analyzed the quantum algorithms identified by the found unitary transformations and generalized the variant models of the original Deutsch's algorithm. We also investigated the relation between the required steps $Q_c$ of the generations to complete the GA process (i.e., the overall run-time) and the mean error $\overline{\epsilon}$ of the found unitaries (i.e., the accuracy). Assuming the typical tendency of the evolutionary methods, we found that $Q_c = O(\alpha e^{-\beta \overline{\epsilon}})$, with $\alpha$ and $\beta$ having finite values.

We expect that our method will be developed further for designing new quantum algorithms or for suppressing the various noises in quantum information processing. We also hope that our method will provide some intuitions or directions in hybridizing machine learning and quantum information science.

%------------------------------
\acknowledgements
%------------------------------

We thank Prof. Jinhyoung Lee and Prof. Hyunseok Jeong. JB thanks Changhyoup Lee and Chang-Woo Lee for helpful discussions. We acknowledge the financial support of the Basic Science Research Program through the National Research Foundation of Korea (NRF) funded by the Ministry of Science, ICT \& Future Planning (No. 2010-0018295 and No. 2010-0015059).

%------------------------------
\appendix
%------------------------------

\section{Proof of Eq.~(\ref{eq:necessary_c})}\label{appendix:variants}

In this appendix, we investigate the necessary condition for the algorithm identified by the found $\hat{U}_1$ and $\hat{U}_2$ to be a variant model of the original Deutsch's algorithm. We start with an arbitrary input $\ket{\psi_\text{in}}$. First, we let 
\begin{eqnarray}
\ket{\psi_1} = \hat{U}_1\ket{\psi_\text{in}}=\alpha\ket{0} + \beta\ket{1},
\end{eqnarray}
where $\ket{0}$ and $\ket{1}$ are defined as the qubit state at the north and the south poles of the Bloch sphere, respectively. The coefficients $\alpha$ and $\beta$ are the complex numbers, satisfying $|\alpha|^2 + |\beta|^2 = 1$. Applying the identified algorithm $\hat{U}_\text{tot}=\hat{U}_3(\mathbf{p}_{3,\text{opt}})\hat{U}_2(x_i)\hat{U}_1(\mathbf{p}_{1,\text{opt}})$ to the input $\ket{\psi_\text{in}}$, we obtain the output states as
\begin{eqnarray}
\ket{\psi_\text{out}(x_i)}=
\left\{
	\begin{array}{ll}
	\hat{U}_3 \left( \alpha \ket{0} + \beta \ket{1} \right) & ~\text{if $x_i$ is constant}, \\
	\hat{U}_3 \left( \alpha \ket{0} - \beta \ket{1} \right) & ~\text{if $x_i$ is balanced},
	\end{array}
\right.
\end{eqnarray}
with the dependence on the given Boolean function $x_i$ ($i=1,2,3,4$). Note that, in order to discriminate the given function $x_i$ by a fixed (von Neumann) measurement, the above two output states should be orthogonal to each other, i.e.,
\begin{eqnarray}
\abs{\left( \alpha^\ast \bra{0} - \beta^\ast \bra{1}\right) \hat{U}_3^\dagger \hat{U}_3 \left( \alpha \ket{0} + \beta \ket{1} \right)}^2 =0.
\label{eq:condi_dis}
\end{eqnarray}
From Eq.~(\ref{eq:condi_dis}), we directly have
\begin{eqnarray}
\abs{\alpha}^2 = \abs{\beta}^2,
\end{eqnarray}
or equivalently,  
\begin{eqnarray}
\ket{\psi_1} = \frac{1}{\sqrt{2}} \left({\ket{0} + e^{i \phi}\ket{1}}\right),
\end{eqnarray}
where $\phi$ is an arbitrary (relative) phase factor. From this proof, we can infer that many generalized versions of the Deutsch's algorithm with numerous sets of $\Theta_j$ and $\mathbf{n}_j$ ($j=1,3$) can give the desired output $\ket{\psi_\text{out}}$, as in Eq.~(\ref{eq:out_st}).

%----------------------------------------------------------------------
% references
%----------------------------------------------------------------------

%\bibliography{common_ga}   % name your BibTeX data bas

\end{document}